\documentstyle[prl,aps,twocolumn]{revtex}

\begin{document}

\title{Pre-Equilibrium Particle Emission and Critical Exponent Analysis}

\author{Wolfgang Bauer and Alexander Botvina\footnote{On
leave from Institute for Nuclear Research, 117312 Moscow, Russia.
Present address: Hahn-Meitner-Institut, 14109 Berlin, Germany.}}

\address{National Superconducting Cyclotron Laboratory, Michigan State
University, East Lansing, MI 48824-1321, USA}

\maketitle

\begin{abstract}
In two different phase transition models of nuclear fragmentation we show
that the emission of pre-equilibrium particles and mixing of events from
different classes cannot be ignored
in the analysis of nuclear fragmentation data in terms of critical exponents,
and we show how the apparent values of the extracted exponents are affected.
\end{abstract}


The premier goal of medium and high energy heavy-ion reactions is the
exploration of the nuclear phase diagram.  On theoretical grounds we expect
infinite nuclear matter to undergo at least two distinct phase transitions.
One is the deconfinement or quark-gluon-plasma phase transition.  The other
is a ``liquid-gas" type phase transition.  It is
believed to be of first order, terminating at the critical point in a second
order transition.  In nuclear multifragmentation reactions one attempts to
map out the liquid-gas coexistence region and locate the critical point.

The first dataset to be interpreted in terms of critical exponents resulted
from proton-induced spectator-fragmentation of krypton and xenon targets
\cite{Min82,Hir84}.  However, it was later shown in the framework of
the percolation model \cite{Bau85} that value of the ``critical"
exponent $\tau$ observed in \cite{Min82,Hir84} was predominantly a result of
mixing of different event classes and integration over impact parameter.
In addition, similar power-law behavior was seen in classical reaction
dynamics simulations, where it was shown to be inconsistent with matter
going through the critical point \cite{Sch87,Lat94,Pra95}.

A significant step forward was then taken by performing event-by-event
analysis of the moments of the mass- or charge-distributions of the fragments
\cite{Cam86,Bau88}.  The result of this analysis of emulsion
data suggested that
nuclei break up similar to percolation clusters \cite{Cam86}.

It was hoped that the problem of impact parameter selection would be less
severe and therefore the critical exponents easier to extract for
participant fragmentation in symmetric heavy ion collisions.  By focussing
on very central collisions as a function of beam energy, a minimum of the
fit parameter $\lambda$, with $\sigma(Z_f) \propto Z_f^{-\lambda}$ was
observed \cite{Li93}.  This confirmed similar observations obtained from
reverse-kinematics reactions \cite{Ogi91}.  It had been predicted that this
minimum would correspond to the actual value of the critical exponent $\tau$
\cite{Bau85}.  However, theoretical calculations showed that there may be
significant topology effects such as the formation of bubbles and toroids
\cite{Bau92,Gro92,Bor93} at work, changing the values of the measured
exponents \cite{Pha93}.

At present, significant effort is also directed at the study of the moments
of the fragment charge distribution in reverse-kinematics reactions,
where the rise and fall of multifragmentation was observed
\cite{Ogi91,Hub91,Bot95}.
Recently the EOS TPC collaboration used the reaction 1 A\,GeV Au + C in an
attempt to reach the critical point of nuclear matter and determine the
critical exponents in the spectator
fragmentation of the residue of the gold nucleus \cite{Gil94,Rit95}.

It is generally agreed that the fragmentation process in proton-induced or
reverse-kinematics reactions proceeds in two steps -- a first pre-equilibrium
step in which
the participants interact and deposit excitation energy into the spectators,
and a second equilibrium step in which the excited (and hopefully
equilibrated!) spectator residue decays.  To this end we have constructed
a hybrid model, in which the pre-equilibrium energy deposition and resulting
residue size are calculated in the framework of an intra-nuclear cascade (INC)
model \cite{BotXX}, and in which we calculate the decay of the residue
within a percolation model \cite{Bau85,Bau88} or statistical multifragmentation
model (SMM) \cite{BotYY} framework.  For any given impact parameter the INC
provides the charge and mass as well as the excitation energy per nucleon
of the spectator residue by calculating a sequence of individual
nucleon-nucleon
collisions and single-particle removals from the nuclear potential well.
It is worth noting that in central collisions the
average residue charge is less than 49, implying the emission of 30
pre-equilibrium charges.

The excitation energy can be converted into a percolation bond-breaking
probability via \cite{Li93}
\begin{equation}
   p_{\rm b} = 1 - \frac{2}{\sqrt{\pi}}  \,
               \Gamma\left(\frac{3}{2},0,\frac{B}{T}\right)\ ,
\end{equation}
where $\Gamma$ is the generalized incomplete gamma function, $B$ is the
binding energy per nucleon in the residue (taken as 6 MeV here), and
$T$ is the temperature as calculated from the INC model, $T=\sqrt{E^*/a}$.
The SMM is microcanonical statistical multifragmentation model where the
decay probabilities into channel $j$ is proportional to the statistical
weight $W_j\propto \exp[S_j(T,A,Z)]$, where $S_j$ is the entropy in channel
$j$.  In the SMM we used a standard set of parameters applied before
for analysis of multifragmentation in peripheral heavy-ion
collisions \cite{Bot95}. In particular the break-up density is
assumed to be 1/3 of the normal nuclear density.

We then arrive at the final inclusive mass distribution (and its moments)
as follows:
For any given event we select an impact parameter with geometrical probability
weighting, calculate $E^*\rightarrow T\rightarrow p_{\rm b}$ and the
pre-equilibrium multiplicity, $m_{\rm pre}$, within the INC stage, and then
use this $p_{\rm b}$ and residue charge, $Z_{\rm res}=79-m_{\rm pre}$,
to obtain the multiplicity, $m_{\rm eq}$, (and charge/mass
distribution) of the equilibrium particles from the percolation stage or
SMM stage.  The total multiplicity is then $m = m_{\rm pre}+m_{\rm eq}$.
A standard integration over impact parameter yields the desired distributions.
Within the assumptions stated above, both hybrid models are then free of
adjustable parameters.

In Fig.\ 1 we compare the results of our models to the data of \cite{Rit95}.
Displayed is the second moment of the charge distribution,
computed event-by-event, and averaged over all events with identical total
charged particle multiplicity,
\begin{equation}
   M_2(m) = \left.
               \sum_i\delta_{m,m_i}
                  \left\{\sum_{Z=1}^{Z_{\rm max}} Z^2\,N_i(Z)\right\}
            \right/
               \sum_i\delta_{m,m_i}\ .
\end{equation}
Here $N_i(Z)$ is the number of fragments of charge $Z$ emitted in event $i$ and
$m_i$ is the total charged particle multiplicity
of this event.  $\sum_i$ is a summation over all events and $\delta_{m,m_i}$ is
the usual Kronecker-$\delta$.  $Z_{\rm max}$ is
the upper cutoff used in the summation over all fragment charges.
Here we should point out the usefulness of displaying
multiplicity-sorted data of $M_2(m)$ by including and excluding the largest,
2$^{\rm}$-largest, 3$^{\rm rd}$-largest, \ldots fragment.  This information
is available in a model-independent way, and it shows the most probable
partition of the system in each multiplicity bin.
In the EOS data \cite{Rit95} displayed here (circles) only part of this
important information is given:
The largest fragment was included for multiplicities
above 26 and excluded below.  For the models we display in the lower curves
$M_2(m)$ excluding the largest fragment and in the higher curves including
the largest fragment for all multiplicities.  Fission events were eliminated
from both the data and the calculations.

We observe an astonishing degree of agreement between the results of the
INC/percolation hybrid (histogram) and the data for all
multiplicities.  We should point out here that the fragmentation of a $Z=79$
residue with the same excitation energy distribution results in average moments
very different from the data, indicating the importance of the emission of
pre-equilibrium charged particles for this observable.  In addition we
performed a calculation fragmenting a $Z=79$ residue at a fixed excitation
energy corresponding to the critical point of the percolation model.  This
resulted in events distributed over the multiplicity interval from 12 to 28,
having approximately constant values of $M_2 \approx 100$ when excluding
and $M_2 \approx 3000$ when including the largest fragment.  This also
indicates the importance of mixing between events of different excitation
energy classes in the observed values of the moments.

The INC/SMM hybrid shows also qualitatively the same features
as the data, but is generally too high on the lower branch and too low on the
upper branch by about a factor of two.  Presumably one could generate better
agreement for the INC/SMM by fitting an excitation energy and residue size
distribution to the data along the method used in \cite{Bot95}. This, however,
is not the goal of the present work.

Rather we would like to point out that the data and
both models have similar slopes of $M_2$ versus $m$
over the entire multiplicity interval.  These slopes were used by the
EOS collaboration to extract the ``critical" exponent $\gamma$.  This is even
more obvious in Fig.~2, where we display $\log M_2(m)$ as a function of
$|m-m_c|$ with $m_c=26$, according to the analysis of \cite{Gil94}.  We see
that data and both models have roughly parallel ``liquid'' and ``gas'' branches
over the interval displayed here.  However, in contradiction to the analysis
of \cite{Ell94}, we do not recover the known critical exponent $\gamma$ (=1.8)
of the percolation model from this figure.  Instead both models yield about
`$\gamma$'=1.4, in accordance with the experimental observation.

We have also performed an analysis of the size of largest fragment as a
function of the total multiplicity.  From an analysis like this, the
authors of ref.\ \cite{Gil94} claim to have extracted the critical exponent
$\beta$.  Recently, one of us \cite{Bau95} pointed out that the presence
of pre-equilibrium particles may change the value of the extracted apparent
exponent.  Within the framework of the present hybrid models we can now
perform a quantitative comparison to the data, as shown in
Fig.~3.  Performing
a linear regression fit to the output of the INC/percolation model
in the interval shown yields
a slope of `$\beta$'=0.35, reasonably close to the value of
0.29$\pm$0.02 extracted from the data, and not close to the nominal value
of $\beta$=0.41 for the percolation model.  A fit to the INC/SMM output
yields `$\beta$'=0.50.

It is dangerous to interpret these exponents extracted in the above way as
the true critical exponents characterizing the universality class of the
actual nuclear phase transition.
Three effects play important roles in this: the extreme finite size of
the fragmenting system, the use of $\log|m-m_c|$ instead of $\log|T-T_c|$
(or $\log|p-p_c|$) for the abscissa, and, most importantly, the mixing of
different event classes of size and excitation in the same multiplicity bin.
This mixing is caused by the use of inclusive, impact parameter integrated
data with different admixtures of pre-equilibrium particles.

The mixture of events with different ``temperature'' into the same multiplicity
bins represents a convolution
\begin{equation}
   O(m) = \int dT\,N(m,T) \otimes O'(T)
\end{equation}
where $O(m)$ and $O'(T)$ is the physical observable as a function of
multiplicity and temperature, and $N(m,T)$ is the (integral normalized) number
distribution function of events with a given temperature and multiplicity.
In Fig.~4, we show $N(m,T)$ as a contour plot for the INC/percolation model.
Each contour level represents a factor of 2 higher value than the one
surrounding it.  One can clearly see that $N(m,T)$ is rather broad, indicating
a large degree of mixing of events with different temperature into the same
multiplicity bin.

While the scaling laws of critical behavior are valid for $O'(T)$ in the
vicinity of the critical point, no such statement can {\em a priori} be made
for the same observable as a function of the multiplicity.  To accomplish
this, the inverse of the convolution function, $N(m,T)^{-1}$
has to be known.  For obvious reasons, the function $N(m,T)$ cannot be
extracted from data in a model-independent way, and therefore the inverse is
also not accessible. For all these reasons, a precise
model-independent determination of critical exponents, as suggested in
refs.\ \cite{Ell94,Gil94,Rit95}, is not correct.

In conclusion, taking
the data of the EOS collaboration we have shown that an interpretation
in terms of critical phenomena should be carried on very carefully.
Mixing of different event classes into the same multiplicity bins and the
contributions of pre-equilibrium particle emission make a
model-independent extraction of critical exponents questionable.

What is possible, however, is a detailed comparison of the data to models
which include pre-equilibrium as well as equilibrium components, and which
emulate as closely as possible the experimental trigger conditions.  We have
used two of the models here to perform this kind of comparison.
We have found, e.g.,
that the INC/percolation model achieves surprising agreement with the
data.  An analysis of the kind employed in \cite{Gil94} for the model output
yields apparent exponents of `$\beta$'=0.35 and `$\gamma$'=1.4,
whereas the known critical exponents for 3d-percolation in the infinite size
limit are $\beta$=0.41 and $\gamma$=1.8. The next step, in our
opinion, could be analysis of exclusive events aimed at finding
actual equilibrated residues. After that we can really investigate
manifestations of the phase transition in the residue-nuclei.  There is
hope that this direction of research will be fruitful.  The rather
surprising agreement of the INC/percolation model (which explicitly contains
a phase transition) with the data given rise to this hope.

\acknowledgements{
This research was supported by NSF grant PHY-9403666 and a Presidential
Faculty Fellow award (W.B.).  One of the authors (A.B.) thanks NSCL/MSU
for hospitality and support.}

\clearpage

\section*{Figure Captions}

\begin{description}
\item[Fig.~1]  Second moment of the charge distribution as a function of the
      charged particle multiplicity for the reaction 1 A\,GeV Au + C.
      Histogram: INC/percolation, line: INC/SMM, circles: data \cite{Rit95}.
\item[Fig.~2]  Second moment of the charge distribution as a function of the
      charged particle multiplicity for the reaction 1 A\,GeV Au + C.
      Histogram: INC/percolation, line: INC/SMM, circles: data \cite{Gil94}.
\item[Fig.~3]  Charge of the largest fragment as a function of the
      charged-particle multiplicity for the reaction 1 A\,GeV Au + C.
      Histogram: INC/perco\-la\-tion, line: INC/SMM, circles: data
\cite{Gil94}.
\item[Fig.~4]  Histogram of the number of events as a function of the
      temperature, $T$, and the total charged particle multiplicity, $m$,
      as predicted by the INC/percolation model.  Each contour line is offset
      by a factor of 2 from the previous one.
\end{description}

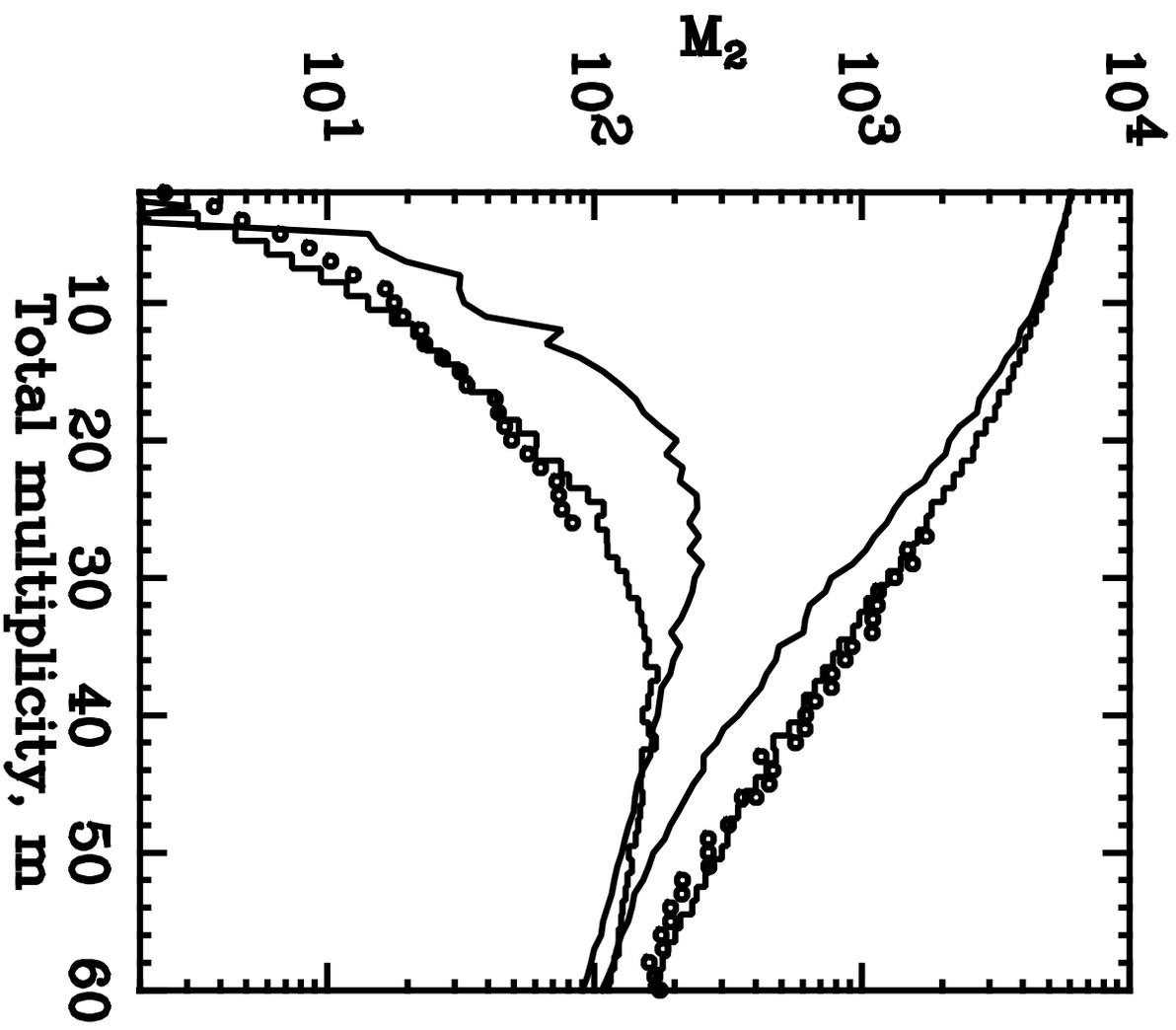
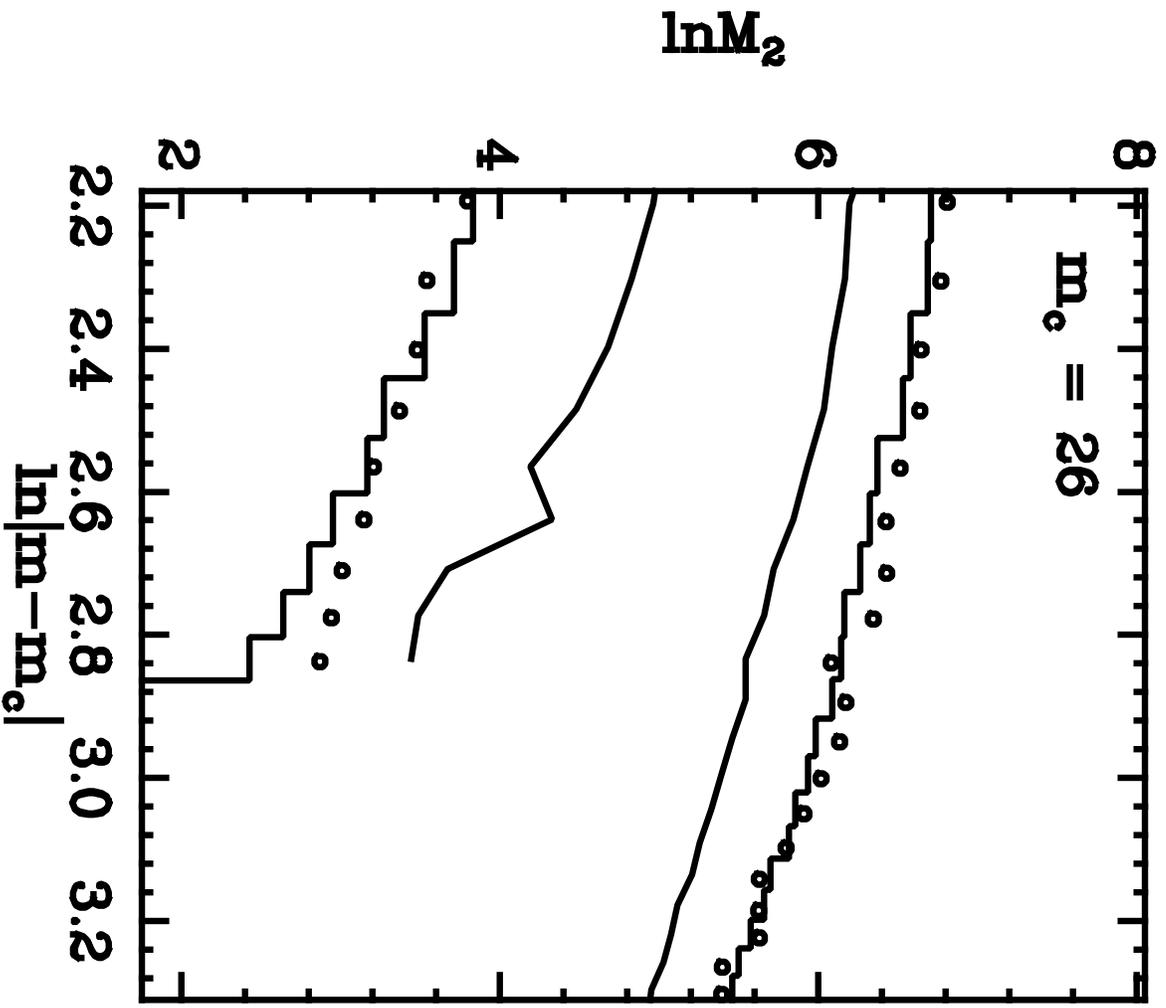
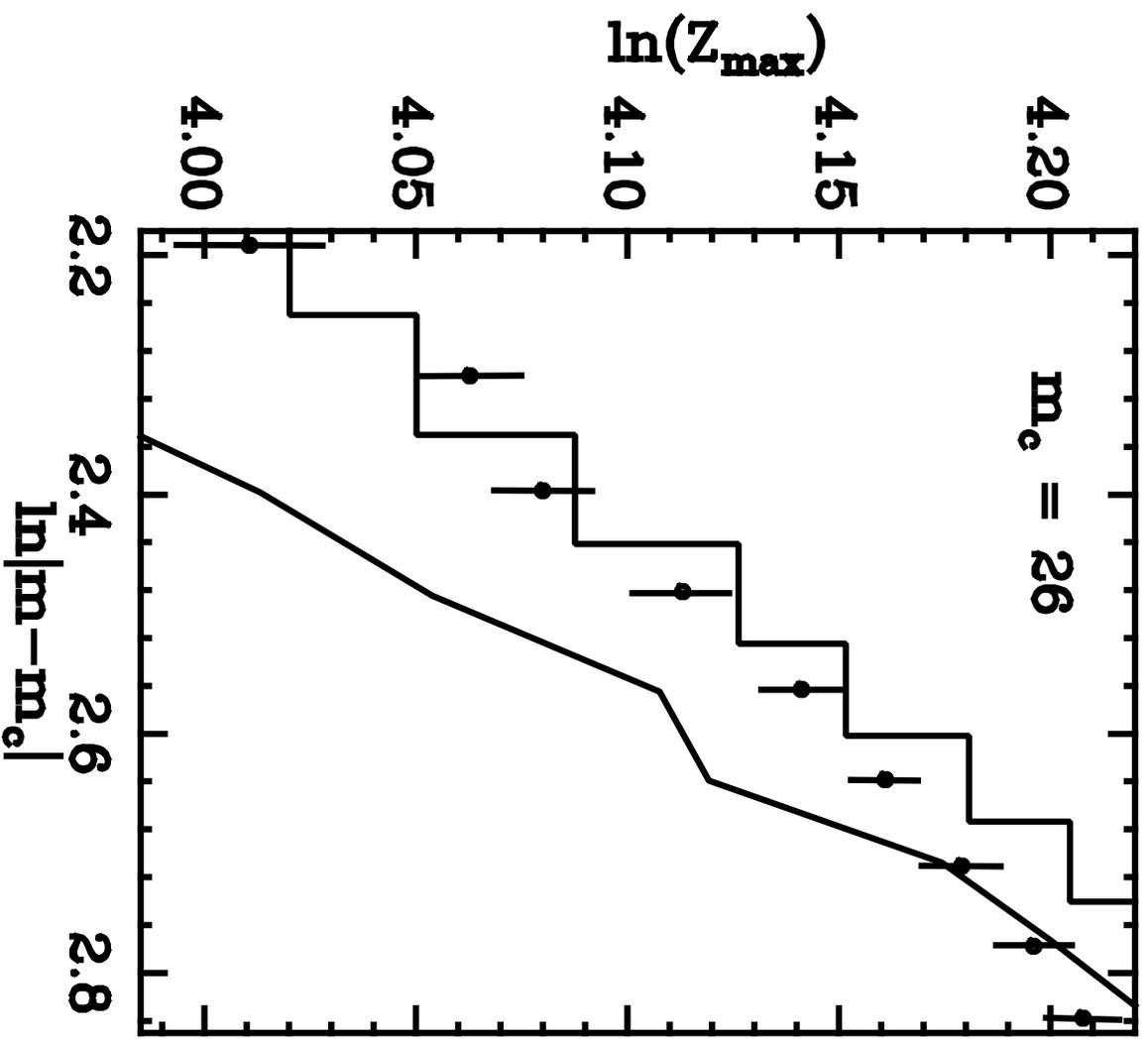
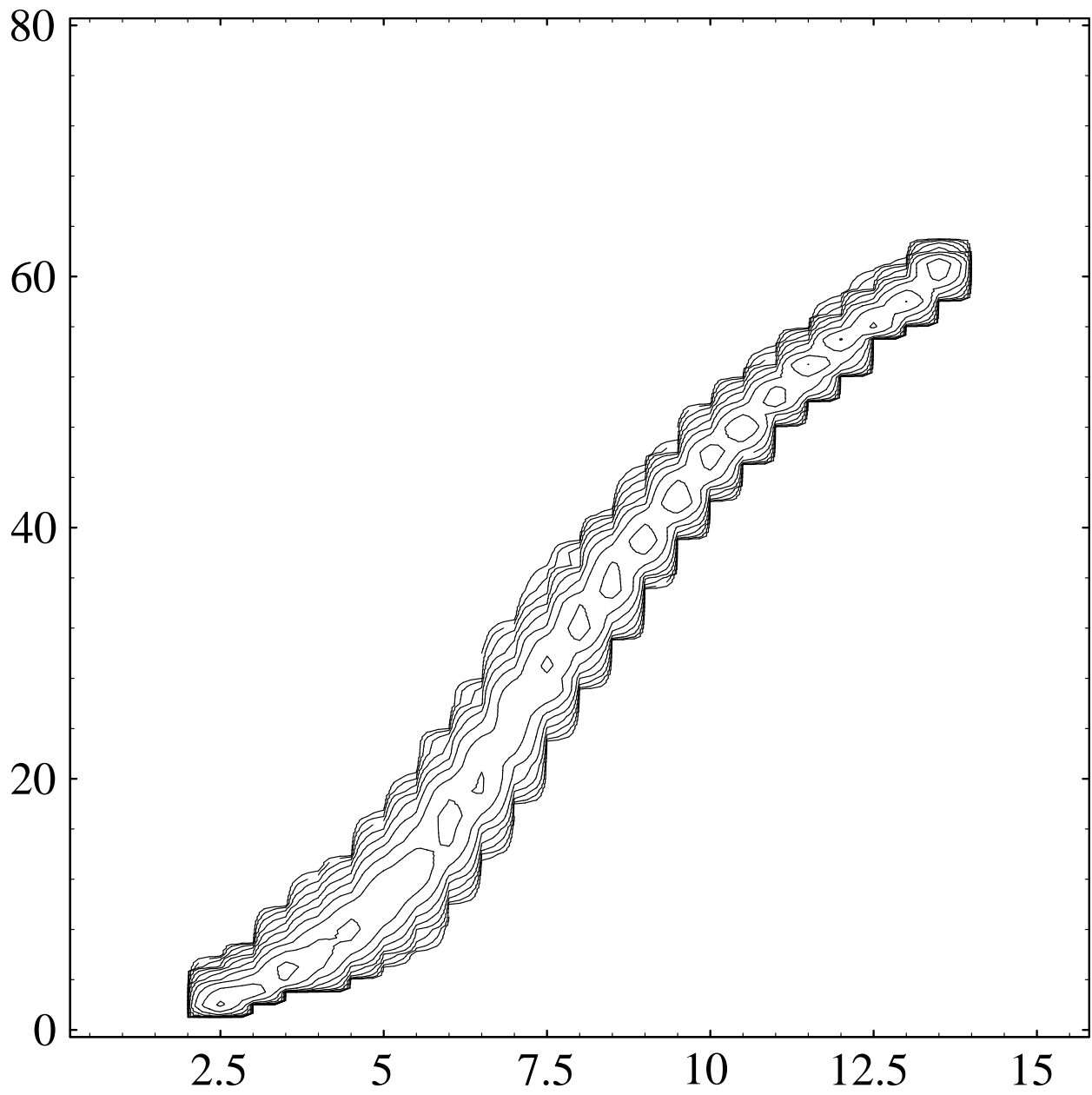
\end{document}